\documentclass[journal=jpcbfk,manuscript=article]{achemso}

\usepackage{amsmath,amssymb}
\usepackage{bm}
\usepackage{graphicx}
\graphicspath{{figures/}}
\usepackage{tikz}
\usetikzlibrary{calc,arrows.meta,decorations.pathmorphing,patterns,positioning}
\usepackage{pgfplots}
\pgfplotsset{compat=1.17}
\newlength{\figw}
\SectionNumbersOn
\usepackage[section]{placeins}
\usepackage[colorlinks=true,allcolors={blue!60!black}]{hyperref}

\author{Pouria A. Mistani}
\email{pouria@examorphic.com}
\affiliation{Examorphic, Inc., Los Gatos, California 95032, United States}


\title{Spatially Resolved Nucleated Polymerization: A Free-Boundary Model of Protein Aggregation in Concentrated Solutions}

\begin{document}

\begin{abstract}
Kinetic models of protein aggregation describe populations by size, not by spatial organization or morphology. We extend Lumry--Eyring nucleated polymerization to a model in which the monomer is a density field and each aggregate is a region bounded by a level set. Growth is a flux condition on the available sites of a surface. Condensation is a reaction between the bonding sites of two surfaces in contact, at a rate set by the bond rate and the contact geometry. The availability of those sites is a field on the interface, and its equilibrium value follows from Wertheim's perturbation theory. The collision efficiency and the Fuchs stability ratio are therefore computed, not fitted. In a well-mixed limit the model's spatial averages satisfy the rate equations term by term; the monomer fraction agrees to eight parts in ten thousand, a difference that arises from equating aggregate size with volume. The condensation kernel's exponent is $0.5806\pm0.0013$ against the $0.600\pm0.010$ fitted to a monoclonal antibody. The computed stability ratio reproduces thirteen of fourteen published conditions at twelve $k_BT$, but only with the bond rate at the top of its range. In a many-body box, aggregates merge at $2.2$ to $3.8$ times the two-body rate.
\end{abstract}

\begin{tocentry}
\includegraphics[width=8.25cm,height=4.45cm,keepaspectratio]{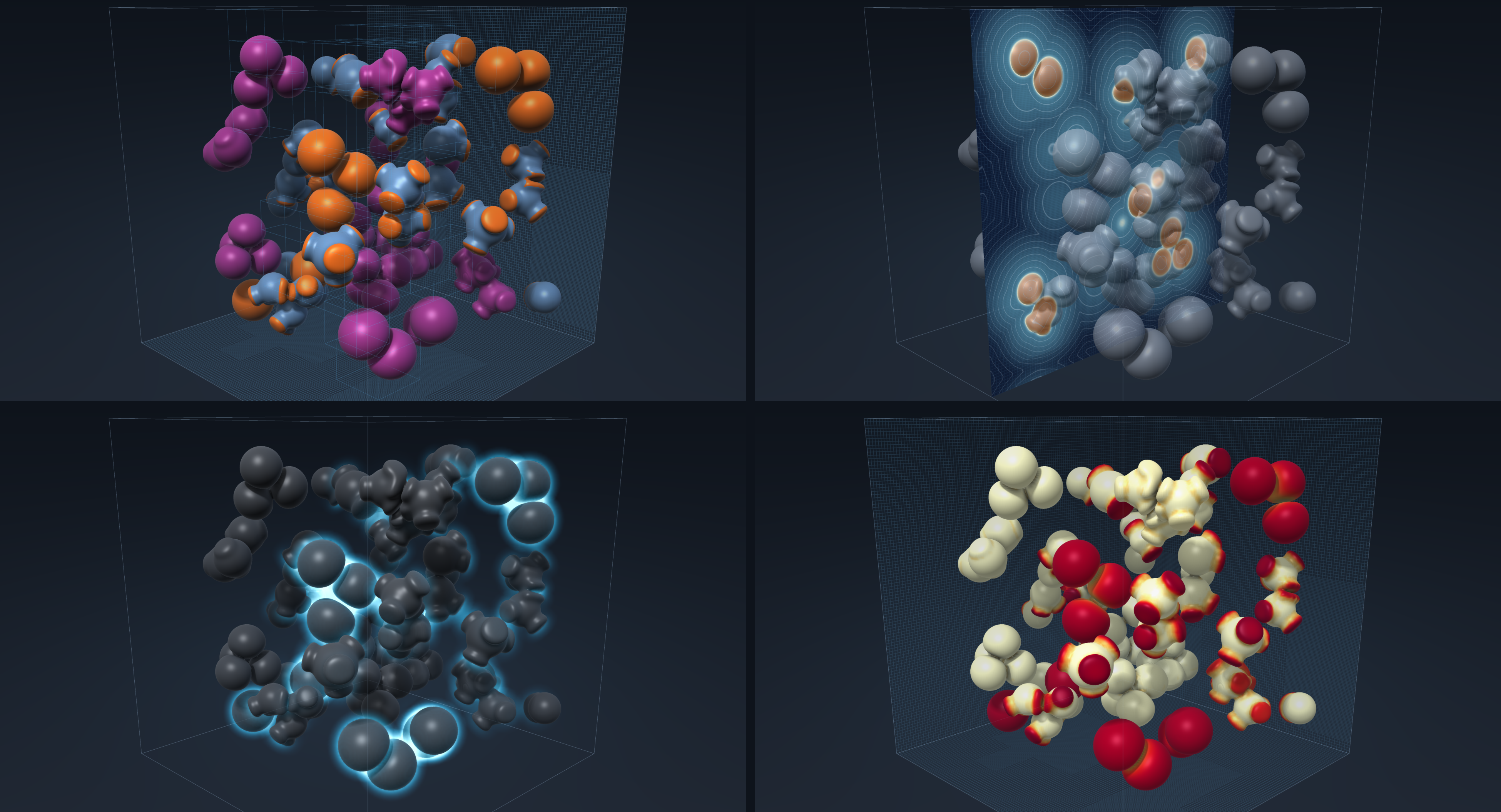}
\end{tocentry}

\section{Introduction}\label{sec:intro}
Monoclonal antibodies are formulated above $100\ \mathrm{mg/mL}$ for subcutaneous delivery,\cite{shire2004challenges} where aggregation limits the concentration a formulation can reach. Aggregates are lost product and a risk factor for immunogenicity.\cite{ROBERTS2014372} The aggregates that one antibody forms under different stresses fall into classes that are distinguished by size, reversibility, conformation and morphology.\cite{JOUBERT201125118,narhi2012classification} A predictive model must therefore describe the population of aggregates together with its spatial arrangement, and not only the distribution of sizes.

Particle-resolved simulation of a concentrated protein solution covers times far shorter than the growth phase, which lasts hours to months.\cite{roberts2007nonnative} The kinetic models used for the growth phase, the Lumry--Eyring nucleated polymerization (LENP) model of Roberts and co-workers\cite{andrews2007a,li2009lumry} and the population balances built on the Smoluchowski equation,\cite{smoluchowski1918versuch,arosio2012population,nicoud2014kinetic} describe the monomer fraction and a size distribution. In those models an aggregate is characterised by its size and at most a fractal dimension. The rate at which two aggregates merge and the fraction of contacts that lead to a merger are supplied as inputs rather than computed. Wertheim's perturbation theory\cite{wertheim1984fluids,fantoni2015wertheim} and coarse-grained simulation\cite{blanco2013,chaudhri2012coarse,skar-gislinge2019a} describe equilibrium structure but not irreversible growth. Cluster--cluster aggregation simulations\cite{meakin1983formation,kolb1983scaling} include a spatial coordinate, but they contain no monomer field, no attachment kinetics and no nucleation. None of these approaches predicts aggregate morphology.

In this paper we add a spatial coordinate to the kinetic description in two steps. The first is a free-boundary formulation. In the protein aggregation model (PAM) introduced here, the monomer is a density field and each aggregate is a region of space. Each nucleation event removes $x$ monomers from the solution and inserts a new region. An aggregate grows through a flux condition imposed on the sticky patches of its surface, and it translates as an overdamped Brownian particle. We represent the regions by the level set method,\cite{osher1988fronts} in which a single scalar field describes the interfaces of the whole population. Two regions that come into contact then merge without an additional rule, as in the island dynamics of epitaxial growth.\cite{CAFLISCH199913,2002PhRvB..65s5403R,mistani2018island}

In the second step we derive the surface rate constant of the growth condition and the collision efficiency, which the first step takes as parameters. The collision efficiency is the fraction of contacts between two aggregates that end in a merger. We treat a merger as a reaction between the bonding sites of two surfaces in contact. Its rate is the intrinsic site--site bond rate multiplied by a kinematic integral over the two level sets, which counts the site pairs within bonding range over the poses in which the surfaces touch. The same integral, evaluated for a monomer at an aggregate surface, determines the coefficient of the growth condition. The fraction of sites on a surface that remain unbonded, or available, is a field on the interface. At long times it relaxes to the value given by Wertheim's theory at the same bond strength that determines the equilibrium monomer density. The collision efficiency and the Fuchs stability ratio therefore follow from the bond rate, the geometry of the two surfaces and the availability of their sites. The derivations are given in the Supporting Information.

\section{Model}\label{sec:model}
\subsection{Fields and transport}
The volume $V_P$ of one protein defines the length $\xi=V_P^{1/3}$, so a protein has unit rescaled volume and a box of side $L_\Omega$ contains $L^3_\Omega$ protein volumes. $\Omega^-$ denotes the aggregate phase, $\Omega^+$ denotes the solution, and $\Gamma_j$ denotes the surface of aggregate $j$ with outward normal $\mathbf{n}$. We describe the free monomer by a number density $\rho(t,\mathbf{x})$ and denote its average over $\Omega^+$ by $<\rho(t)>$. The LENP monomer fraction is then $m(t)=\,<\rho(t)>/<\rho(0)>$. Because each protein has unit rescaled volume, $<\rho(0)>$ is also the packing fraction of the protein. For an immunoglobulin of hydrodynamic diameter $\sigma=10.2\ \mathrm{nm}$ and molar mass $150\ \mathrm{kDa}$,\cite{salinas2010understanding} $\xi\approx8.2\ \mathrm{nm}$, and $<\rho(0)>$ is $0.22$ at $100\ \mathrm{mg/mL}$ and is $0.45$ at $200\ \mathrm{mg/mL}$. Because these values are not small, corrections of relative order $<\rho(0)>$ are retained throughout.

On $\Omega^+$ the monomer diffuses at the rescaled hopping rate $D$ and is depleted by nucleation, which consumes $x$ monomers per event,
\begin{align}
    \partial_t\rho &= \nabla\cdot(D\nabla\rho) - x\,\frac{dn_x}{dt}, \qquad
    \frac{dn_x}{dt} = \frac{1}{\tau_n}\frac{<\rho^x>}{<\rho(0)>^{x-1}} \label{eq::transport}
\end{align}
where $n_x$ is the accumulated number of events per unit volume of solution and $\tau_n$ is the nucleation timescale, which sets the rescaled time $\theta=t/\tau_n$. Each time the accumulated number of events in the solution, $n_x|\Omega^+|$, passes an integer, we insert a new nucleus as a ball of rescaled volume $x$. The ball is placed at a positive distance from every interface, either at a random site or at the point where $\rho^x$ is largest. Inside an aggregate the bound protein obstructs the transport of interstitial monomer. We set the interior diffusivity $D^-$ to zero, so no flux reaches the interface from the interior.

\subsection{Growth at the interface}
An aggregate grows by capturing monomers at its surface in packets of $\delta$, and the diffusive flux of monomer toward the surface supplies them. A protein surface is heterogeneous,\cite{blanco2013} and the antibody considered here has two sticky patches. The growth condition is therefore imposed only on the sticky part $\Gamma^s$ of the interface of a soluble aggregate,
\begin{align}
    \mathbf{x}\in\Gamma^s_j:\quad -D\nabla\rho\cdot\mathbf{n}\big|_+=\frac{\delta}{\tau_g}\frac{\rho^\delta_{eq}-\rho^\delta}{<\rho(0)>^{\delta-1}} \label{eq::robin}
\end{align}
where $|_+$ is the trace from the solution side and $\rho_{eq}$ is the equilibrium monomer density given by Wertheim's theory in section~\ref{subsec:closure}. The surface time constant $\tau_g$ is distinct from the growth timescale $\tau^{\mathrm{LENP}}_g$ of the rate equations. On the non-growing remainder $\Gamma^n$ of the interface, the normal flux vanishes from both sides.

The partition of $\Gamma$ into $\Gamma^s$ and $\Gamma^n$ is defined by a second field $\psi^s$, with $\Gamma^s=\Gamma\cap\{\psi^s<0\}$, which moves with the aggregate. The sticky area $\mathcal{A}_i$ of an aggregate of size $i$ is therefore computed and not prescribed. On a ball-shaped nucleus each patch is the part of the surface inside a cone from the centre. A compact aggregate growing radially therefore keeps the same sticky fraction at every size, and its sticky area scales as $\mathcal{A}_i\propto i^{\alpha_\mathcal{A}}$ with $\alpha_\mathcal{A}=2/3$. If instead a patch of fixed cross-section is carried outward, $\alpha_\mathcal{A}=0$, which is the end growth of a chain for which the rate equations are written (Supporting Information, section S3.2.1). Above a critical size $n^\ast$ an aggregate becomes insoluble and its whole interface switches to the non-growing condition.

Each bound monomer adds one unit of rescaled volume, so the growth contribution to the normal velocity is the sum of the fluxes incident from the two sides,
\begin{align}
    v^{\mathrm g}_n=D\nabla\rho\cdot\mathbf{n}\big|_+-D^-\nabla\rho\cdot\mathbf{n}\big|_- \label{eq::vel}
\end{align}
which is the exact interfacial balance at every point. As the interface advances, it also encloses the free monomer present in the solution it sweeps, and that monomer enters the aggregate without crossing $\Gamma$ as a flux. The number of monomers an aggregate contains and the volume it occupies therefore differ at relative order $<\rho>$.

\subsection{Translation and the dimensionless groups}
Aggregate $j$ occupies the connected region $\Omega^-_j$, whose volume defines its size, and its radius of gyration $R_{g,j}$ follows from the moment tensor of that region. We take the hydrodynamic radius to be $\tilde{R}_j=\sqrt{5/3}\,R_{g,j}$, the ratio for a uniform sphere, which we apply at every morphology. It is exact for a ball and an overestimate for an open aggregate. The flow about a moving aggregate is a Stokes flow, so its mobility is $b_j=1/6\pi\eta_sR_j$, where $\eta_s$ is the solvent viscosity and $R_j=\xi\tilde{R}_j$ is the hydrodynamic radius in physical units. In rescaled variables the diffusivity and the Brownian velocity of aggregate $j$ are
\begin{align}
    \tilde{D}^{\mathrm a}_j=\frac{k_BT\tau_n}{6\pi\eta_s\xi^3}\frac{1}{\tilde{R}_j},\qquad
    \tau_n\mathbf{U}_j=\sqrt{2\tilde{D}^{\mathrm a}_j}\,\boldsymbol{\zeta}_j(\theta) \label{eq::langevin}
\end{align}
where $\boldsymbol{\zeta}_j$ is a Gaussian white noise. A point of $\Gamma_j$ therefore moves at the normal velocity $v_n=v^{\mathrm g}_n+\mathbf{U}_j\cdot\mathbf{n}$. The model includes no force between aggregates and no rotation.

The rate equations contain the groups $\beta_{gn}=\tau_n/\tau^{\mathrm{LENP}}_g$ and $\beta_{cg}=\tau^{\mathrm{LENP}}_g/\tau_c$, which compare the nucleation, growth and condensation timescales. Translation introduces one further group, $G=k_BT\tau_n/6\pi\eta_s\xi^3$, which sets the aggregate diffusivity in rescaled units. For the immunoglobulin above $G$ is of order $10^{9}$ to $10^{12}$. The computations that include the monomer field were run at $G=1$ and $10$, and those with rigid aggregates were run at $G$ between $4$ and $256$. In these computations aggregates therefore move far more slowly relative to the monomer than in a formulation, and in the rate equations this slower transport appears as a smaller $\beta_{cg}$. At fixed $\theta$ the computed populations condense as a solution with a stability ratio of order $10^{5}$ to $10^{6}$ would, and every merger rate reported scales with $G$. At a single aggregate, growth and diffusion compete through the Damk\"ohler number $\mathrm{Da}_j=\delta m^{\delta-1}\tilde{R}_j/\tau_gD$. The reduction derived below holds in the well-mixed limit, in which diffusion across the box is faster than nucleation, growth and condensation. This limit is $\varepsilon=L^2_\Omega\max\{1,\beta_{gn},\beta_{gn}\beta_{cg}\}/D\tau_n\ll1$, together with a small aggregate volume fraction $\phi=|\Omega^-|/L^3_\Omega\ll1$ and uncorrelated aggregate positions.

\subsection{The equilibrium closure}\label{subsec:closure}
Wertheim's first-order perturbation theory\cite{wertheim1984fluids,fantoni2015wertheim} for hard spheres carrying $M$ square-well sites gives the fraction $X$ of unbonded sites through a mass-action law in the packing fraction $\eta$, the well depth $\epsilon_W$ and the range $a_W$ (Supporting Information, section S5). At first order the sites bond independently, so the density of proteins carrying no bond is $\rho_WX^M$, which we identify with the equilibrium value of the monomer field,
\begin{align}
    \rho_{eq}=\rho_WX^M,\qquad \eta=\xi^3\rho_W=\,<\rho(0)>,\qquad m_{eq}=X^M \label{eq::rho_eq}
\end{align}
At this point the identification is an assumption. It follows as a consequence once detailed balance is imposed on the contact kinetics below. When $\rho=\rho_{eq}$ at the interface, the growth flux vanishes and $v^{\mathrm g}_n=0$, so monomer depletion stops at $m_{eq}$ rather than proceeding to zero. We verified this limit in a closed box that initially contains aggregates and in which no nucleation occurs. At a packing fraction of $0.05$ the monomer fraction relaxes to within $3.3\times10^{-6}$ of $m_{eq}$ (Supporting Information, section S8.14).

\subsection{Contact kinetics}
In the model as stated so far, the surface time constant $\tau_g$ is a parameter. Because two regions merge on first contact, the collision efficiency is also fixed at unity. We now replace both by a reaction between the bonding sites of two surfaces in contact. A protein of radius $R_1$ has $M$ such sites at the areal density $n_s=M/4\pi R^2_1$, of which a fraction $f$ is available. Two free sites within the square-well width $a_W$ of each other bond at the intrinsic rate $k_{\mathrm{on}}$. A bond breaks at the rate $k_{\mathrm{off}}=k_{\mathrm{on}}e^{-\beta\epsilon_W}$, which detailed balance fixes, so bond breaking requires no input beyond the bond strength $\epsilon_W$ already used in the equilibrium closure. A monomer bonds only in its reactive conformation, which is present at the fraction $\chi$. A bond between two aggregate surfaces becomes a permanent merger with probability $\eta_c$. The kinetic inputs of the model are therefore $\chi$, $\eta_c$ and $k_{\mathrm{on}}$.

Let $\Pi_{i,j}(\mathsf{g})$ be the number of pairs of available sites within $a_W$ of each other, one on each surface, when a rigid motion $\mathsf{g}$, an element of the group $SE(3)$ of rotations and translations, places body $j$ against body $i$, where a body is an aggregate or a monomer. The contact functional is the equilibrium average of that count over the poses at which the two regions touch without overlapping,
\begin{align}
    \mathcal{C}_{i,j}=g_{i,j}\int_{SE(3)}d\mathsf{g}\;\mathbf{1}\big(\Omega^-_i\cap\mathsf{g}\Omega^-_j=\varnothing\big)\,\Pi_{i,j}(\mathsf{g}) \label{eq::contactfunctional}
\end{align}
where $g_{i,j}$ is the contact value of the pair correlation of aggregate positions. The functional $\mathcal{C}_{i,j}$ generalises the two-body bonding volume of Wertheim's theory from two spheres to two regions of arbitrary shape. Each surface integral in $\Pi_{i,j}$ is an integral over a level set, so $\mathcal{C}_{i,j}$ is computed directly from the two level sets (Supporting Information, section S3.4.2).

Aggregate $i$ captures monomer at the rate $k_{\mathrm{on}}\chi\,\mathcal{C}_{1,i}\,\rho|_+$, where $\mathcal{C}_{1,i}$ is the contact functional of a monomer and aggregate $i$. Dividing this rate by the available area of the aggregate gives the flux of equation \ref{eq::robin}, with its coefficient now derived,
\begin{align}
    \frac{1}{\tau_g(\mathbf{x})}=\frac{1}{\tau^\infty_g}\sqrt{\big(1+R_1\varkappa_1\big)\big(1+R_1\varkappa_2\big)},\qquad
    \frac{1}{\tau^\infty_g}=\frac{\pi^2}{2}\,k_{\mathrm{on}}\,\chi\,g_1\,a^4_W\,n^2_s\,R_1 \label{eq::taugcurv}
\end{align}
in which $\varkappa_1$ and $\varkappa_2$ are the principal curvatures of $\Gamma$, filtered at the scale below which the model represents no structure, and $g_1$ is the contact value of the pair correlation of a monomer at an aggregate surface (Supporting Information, section S3.4.3). The parameter $\tau_g$ is therefore replaced by the product $k_{\mathrm{on}}\chi$.

Two aggregates $i$ and $j$ merge at the rate $\mathcal{K}^{\mathrm R}_{i,j}=k_{\mathrm{on}}\eta_c\mathcal{C}_{i,j}$. We eliminate $k_{\mathrm{on}}$ between this rate and equation \ref{eq::taugcurv}. The bonding range $a_W$ and the site density $n_s$ then cancel, and the result gives the condensation kernels of the rate equations and the group $\beta_{cg}$,
\begin{align}
    \kappa_{i,j}=\frac{f_if_j\big(R_i+R_j\big)R_iR_j}{\big(1+\delta^{\mathrm K}_{ij}\big)f^2_xR^3_x},\qquad
    \beta_{cg}=\frac{\eta_c}{\chi}\frac{f_xR_x}{R_1} \label{eq::kappacontact}
\end{align}
where $R_x$ and $f_x$ are the radius and the availability of a nucleus. This is the reaction-limited kernel, which applies when diffusion brings pairs together much faster than they react, and it contains no fitted parameter. When transport is not fast, the reaction acts in series with the transport-limited rate coefficient $\mathcal{K}^{\mathrm S}_{i,j}$ derived in section~\ref{sec:reduction}, and the two combine as
\begin{align}
    \frac{1}{\mathcal{K}_{i,j}}=\frac{1}{\mathcal{K}^{\mathrm S}_{i,j}}+\frac{1}{\mathcal{K}^{\mathrm R}_{i,j}},\qquad
    W_{i,j}=1+\frac{\mathcal{K}^{\mathrm S}_{i,j}}{\mathcal{K}^{\mathrm R}_{i,j}} \label{eq::collinskimball}
\end{align}
which is the combination of Collins and Kimball.\cite{collins1949diffusion} The collision efficiency of the model is $1/W_{i,j}$, the ratio of the merger rate to the transport-limited rate. $W_{i,j}$ is the Fuchs stability ratio,\cite{fuchs1934stabilitat} which is a fitted parameter in a population balance and is computed here.

\subsection{Two ways of integrating the model}\label{subsec:branches}
Merging requires two aggregates to be in contact, and whether the positions of the aggregates must be followed for that purpose depends on which step limits the merger rate. If transport is fast compared with bonding (the reaction-limited case, $W_{i,j}\gg1$), every pair has time to sample all relative positions before it bonds, and the merger rate depends only on the two aggregates themselves, through the contact functional of equation \ref{eq::contactfunctional}. If bonding is fast compared with transport (the transport-limited case, $W_{i,j}\approx1$), the merger rate depends on how often the aggregates actually meet, and their positions and motion must be followed. We therefore integrate the model in two ways, which we call the mean-field branch and the resolved branch.

In the mean-field branch the aggregates do not diffuse. An aggregate is moved when it takes part in a merger, and also when it comes into contact with another aggregate without a merger having been chosen, so that the two remain separate. The monomer field is still solved in space and every aggregate grows from it, but mergers are not detected geometrically. Instead, in each time step the mergers are chosen at random among all pairs, with each pair merging at the rate $\mathcal{K}^{\mathrm R}_{i,j}$, as in a population balance. The two aggregates of each chosen pair are then placed in contact, at the relative position and orientation that maximises $\Pi_{i,j}$ without overlap, and they form one aggregate. Because the merger rate of a pair does not depend on where the other aggregates are, this branch contains no spatial correlation between aggregates.

In the resolved branch each aggregate moves by Brownian translation according to equation \ref{eq::langevin}, and a merger occurs only when two level sets actually come into contact. A pair whose surfaces are within $a_W$ of each other bonds at the surface reactivity $k_r=k_{\mathrm{on}}\eta_c\int^{a_W}_0\Pi_{i,j}(s)\,ds$, which is imposed as a radiation boundary condition on the Brownian step.\cite{erban2007reactive,erban2009stochastic} A pair that does not bond is reflected, so that two aggregates cannot overlap. This branch therefore includes the effect of the positions of all other aggregates on the rate at which a given pair meets. In the computations reported here the aggregates of the resolved branch are rigid and do not grow.

The two branches agree where both are valid, and equation \ref{eq::collinskimball} gives the form of that agreement. Section~\ref{subsec:twobranches} compares them in populations of many aggregates.

\subsection{Site availability}\label{subsec:availability}
We call a site buried once it has bonded to a neighbouring protein. Burial is then a state of the sites and not a property of the shape of the aggregate. We define the available fraction $f(t,\mathbf{x})\in[0,1]$ as a bounded field on the sticky part of the interface, and the available area of aggregate $i$ is $\mathcal{A}_i=\int_{\Gamma^s_i}f\,dA$. Let $f_{\mathrm{out}}$ be the fraction of surface sites that face the solution and have no bonding partner. Let $X_s$ be the unbonded fraction of the remaining sites, so that $f=f_{\mathrm{out}}+(1-f_{\mathrm{out}})X_s$. Following a point of the interface, $X_s$ obeys
\begin{align}
    \frac{dX_s}{dt}\bigg\vert_\Gamma=-k_2X^2_s+k_{\mathrm{off}}\big(1-X_s\big)+\omega\big(X_{s,\mathrm{att}}-X_s\big) \label{eq::availevolution}
\end{align}
Here $k_2=k_{\mathrm{on}}\mathcal{B}n_s(1-f_{\mathrm{out}})$ is the rate at which unbonded sites pair within the surface layer, and $\mathcal{B}$ is the surface bonding area. The quantity $X_{s,\mathrm{att}}=1-z_g/[M(1-f_{\mathrm{out}})]$ is the unbonded fraction of freshly grown surface, which is below one because an attaching protein bonds $z_g$ of its sites inward to the aggregate. The first term describes bond formation, and the second term describes bond breaking. The third term describes the creation of fresh surface by growth, which adds $\omega=v_n/\ell_1$ monolayers per unit time, where $\ell_1=4R_1/3$ is the thickness of one monolayer.

With no growth the fixed point of equation \ref{eq::availevolution} is
\begin{align}
    f_{eq}=f_{\mathrm{out}}+\big(1-f_{\mathrm{out}}\big)X_{s,eq},\qquad
    X_{s,eq}=\frac{2}{1+\sqrt{1+4n_s\big(1-f_{\mathrm{out}}\big)\mathcal{B}e^{\beta\epsilon_W}}} \label{eq::availeq}
\end{align}
which is the mass-action law behind equation \ref{eq::rho_eq} applied to the sites of the surface. Because bonds break as well as form, burial saturates at $f_{eq}$ instead of proceeding to zero. The same bond strength $\epsilon_W$ determines both $\rho_{eq}$ and $f_{eq}$. We call a surface aged when its availability has reached the equilibrium value $f_{eq}$. On a growing surface this requires that fresh surface be created much more slowly than bonds break, $\omega\ll k_{\mathrm{off}}$. That condition is stronger than the condition that the field relax quickly (Supporting Information, section S14).

At a merger the available sites in the two contact zones bond across the gap one to one. The region that lies within bonding range of both surfaces is added to the merged aggregate as a band of volume around the neck, and the availability of this band is zero (Supporting Information, section S15).

Equation \ref{eq::availevolution} adds no kinetic parameter. The four geometric closures that enter it are the coordination $z_s=4$, the outward-facing fraction $f_{\mathrm{out}}=0$, the number $z_g=1$ of sites that an attaching protein bonds inward, and the availability $f_x=0.5$ of a fresh nucleus. The plausible range of each is given in the Supporting Information, section S3.5.5. We use the bond strength $\beta\epsilon_W=12.0$, which gives $f_{eq}=0.0121$, in the mean-field computation of the size distribution and the mergers, and in the computations of the resolved branch. The two computations of section~\ref{subsec:kernelexponent} that include the field use $\beta\epsilon_W=10.0858$, which gives $f_{eq}=1/32$.

Suppose that in a population the hydrodynamic radius scales with size as $\tilde{R}_i\propto i^{1/d_R}$, where $d_R$ is the mass--radius exponent, and that the availability scales as $f_i\propto i^{\alpha_f}$. The reaction-limited kernel is then homogeneous of degree $2\lambda_\kappa$ in the sizes, with
\begin{align}
    \lambda_\kappa=\alpha_f+\frac{3}{2d_R} \label{eq::lambdakernel}
\end{align}
In the aged limit the availability of every surface is $f_{eq}$, independent of size. The availability exponent $\alpha_f$ then vanishes, and $\lambda_\kappa$ takes its saturated value $3/(2d_R)$, which depends only on the geometry of the population. Away from that limit $\alpha_f$ is negative, because the availability of an aggregate depends on its age and age correlates with size. A surface flow of singly bonded proteins could compact a merger product, but such a flow is frozen at these bond strengths. In the model a product therefore becomes more compact only through the growth that follows the merger. The exponent $d_R$ consequently depends on the relative rates of growth and condensation as well as on the model (Supporting Information, section S3.5.7).

\begin{figure}[!htbp]
\centering
\includegraphics[width=0.88\linewidth]{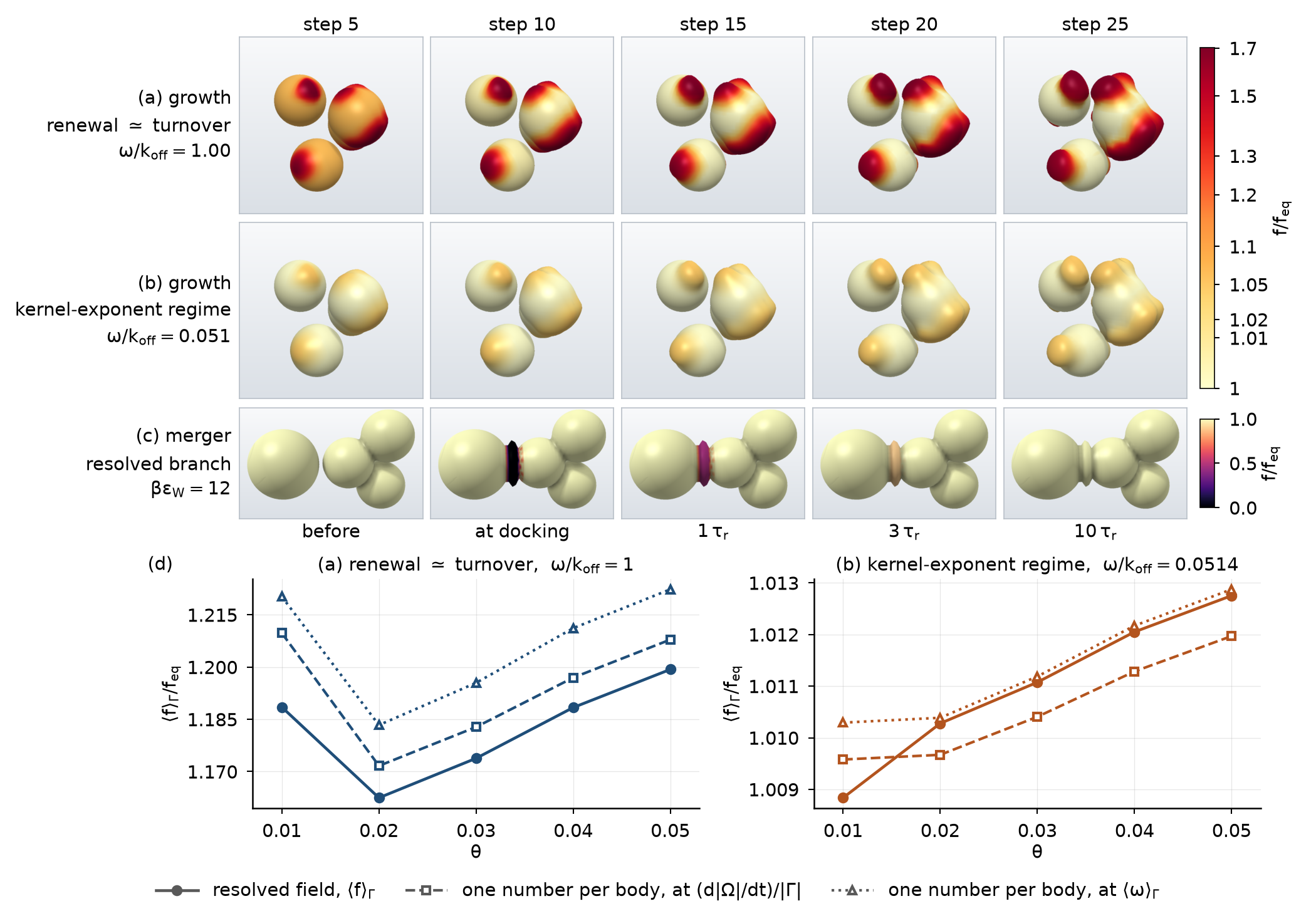}
\caption{\label{fig::availfield}The site-availability field $f/f_{eq}$ on aggregate surfaces. (a) and (b) The field on the growing surfaces of the patched computation at $\mathrm{Da}=4$ (section~\ref{subsec:morphology}), with surface creation (a) as fast as unbonding and (b) at the ratio of surface creation to unbonding of the kernel-exponent computation. The light end of the colour scale is $f_{eq}$. (c) A merger of two rigid aggregates in the resolved branch, over ten relaxation times after contact. (d) The surface mean of the field compared with the single value per aggregate of the mean-field branch.}
\end{figure}

Figure \ref{fig::availfield} shows the field on a surface. No population computation below evaluates the field point by point on a growing surface. In the mean-field branch each aggregate has a single availability, which is advanced at the mean growth speed of that aggregate. In the resolved branch the aggregates do not grow. To show the field on a growing surface, we integrated equation \ref{eq::availevolution} on the surfaces recorded during the computation with sticky patches at $\mathrm{Da}=4$ of section~\ref{subsec:morphology}, which gives rows (a) and (b). The rate $\omega$ of surface creation was taken from the recorded interface velocity. The field was computed from the recorded growth and did not act back on it. The growth coefficient of that computation did not contain $f$, so no growth rate or exponent is obtained from these rows. When the rate $\omega$ of surface creation equals the unbonding rate $k_{\mathrm{off}}$, the median of $f/f_{eq}$ at the last surface is $1.55$ on the sticky part of the interface, and it is $1.00$ on the non-growing part. At the ratio $\omega/k_{\mathrm{off}}$ of the kernel-exponent computation, the same medians are $1.036$ and $1.000$. The surface is then aged, as the saturated exponent $3/(2d_R)$ assumes. In a merger of two aged rigid aggregates (row c) the contact zones and the band around the neck are depleted, and the availability on the surface of that band then recovers from $0.076$ to $0.91$ of $f_{eq}$ within three relaxation times. At the ratio $\omega/k_{\mathrm{off}}$ of the kernel-exponent computation, the single availability per aggregate used by the mean-field branch differs from the surface mean of the field by at most $0.144$ per cent (panel d; Supporting Information, section S3.5.7).

\subsection{Morphology and its stability}
Morphology is quantified by the moment tensor of each region, whose equation of motion is driven by the growth velocity weighted by the position at which capture occurs (Supporting Information, section S6.2). The model contains no surface energy. Without one, a diffusion-limited moving boundary held at a fixed density would ordinarily be unstable at every wavelength.\cite{mullins1963morphological} The growth condition here instead contains a finite surface rate constant. We linearised equations \ref{eq::robin} and \ref{eq::vel} about a growing ball that is sticky over its whole surface. Every shape perturbation of spherical-harmonic degree $\mathcal{L}\ge2$ then decays relative to the radius if and only if $\delta\,\mathrm{Da}_j\le1$. Because $\mathrm{Da}_j$ increases with the radius, this is a condition on aggregate size. At the timescales of a formulation it is satisfied with a margin of one order of magnitude, even where aggregates are largest and growth is fastest. The result applies to a uniformly sticky interface. At the boundary of a patch the interface develops a crease, and because the model does not specify how that crease evolves, its evolution is determined by the numerical representation. The linearisation and the crossover wavelength below which the curvature dependence of equation \ref{eq::taugcurv} roughens an interface are given in the Supporting Information, sections S6.5 and S3.4.4.

\section{Reduction to nucleated polymerization}\label{sec:reduction}
The LENP equations\cite{andrews2007a,li2009lumry} describe the monomer fraction $m$ and the fractions $a_j$ of soluble aggregates of size $x\le j<n^\ast$, normalised on $<\rho(0)>$,
\begin{align}
    \frac{dm}{d\theta} &= -xm^x-\delta\beta_{gn}m^\delta\lambda_0 \nonumber\\
    \frac{da_x}{d\theta} &= m^x-\beta_{gn}a_xm^\delta-\beta_{cg}\beta_{gn}a_x\Big(\kappa_{x,x}a_x+\sum_{j}\kappa_{x,j}a_j\Big) \label{eq::LENP_ax}\\
    \frac{da_i}{d\theta} &= \beta_{gn}(a_{i-\delta}-a_i)m^\delta-\beta_{cg}\beta_{gn}a_i\Big(\kappa_{i,i}a_i+\sum_j\kappa_{i,j}a_j\Big)+\beta_{cg}\beta_{gn}\sum_{j\le i/2}\kappa_{i-j,j}a_{i-j}a_j \label{eq::LENP_aj}
\end{align}
where $\lambda_0=\sum_ja_j$ is the number of aggregates per unit volume and the kernels $\kappa_{i,j}$ are free parameters. To compare PAM with these equations, we take $m$ as the spatial average $m=\,<\rho>/<\rho(0)>$ and $a_i$ as the number of connected aggregate regions of volume $i$ per unit volume of the box.

\emph{Statement.} Consider PAM in the well-mixed limit $\varepsilon\rightarrow0$, with $\phi\ll1$ and uncorrelated positions. Let $D^-=0$, $\rho_{eq}=0$ and $n^\ast\rightarrow\infty$, and let capture be resolved as a jump process in packets of $\delta$ monomers. Suppose further that the sticky area is the same at every soluble size. Then the spatial averages of PAM satisfy equations \ref{eq::LENP_ax} and \ref{eq::LENP_aj}. Every equality holds to relative order $\varepsilon$, except for a difference of relative order $<\rho(0)>$ that remains as $\varepsilon\rightarrow0$ and arises from identifying the size of an aggregate with its volume. The timescales of the rate equations are
\begin{align}
    \tau^{\mathrm{LENP}}_g=\frac{\tau_g}{<\rho(0)>\mathcal{A}},\qquad \tau_c=\frac{2}{<\rho(0)>\mathcal{K}_{x,x}} \label{eq::taug}
\end{align}
and the kernels are supplied by the transport problem below. The proof compares the two descriptions term by term through the rate equations (Supporting Information, section S4). The hypothesis on the sticky area holds for end growth, $\alpha_\mathcal{A}=0$, and fails for the cone patches, for which $\alpha_\mathcal{A}=2/3$ on a compact aggregate. The departure of PAM from the reduced equations is therefore controlled by the sticky-area exponent $\alpha_\mathcal{A}$.

Two aggregates collide when the relative diffusion of their centres brings them to the contact radius. Treating contact as absorption at that radius gives the Smoluchowski rate coefficient $\mathcal{K}^{\mathrm S}_{i,j}=4\pi(\tilde{R}_i+\tilde{R}_j)(D^{\mathrm a}_i+D^{\mathrm a}_j)/\xi^2$. Each diffusivity is inversely proportional to the radius, so the absolute size cancels between the contact radius and the mobilities, and morphology enters only through the ratio of the radii. The kernels of the rate equations are then
\begin{align}
    \kappa_{i,j}=\frac{1}{1+\delta^{\mathrm K}_{ij}}\frac{\big(\tilde{R}_i+\tilde{R}_j\big)^2}{2\tilde{R}_i\tilde{R}_j} \label{eq::kappa_derived}
\end{align}
with no fitted parameter. This kernel has its minimum at equal radii and grows without bound as the radii become unequal. It is therefore not the size-independent constant that a fitted kernel is usually taken to be. The single-absorber solution behind it neglects the other aggregates, which also absorb partners, and it requires an aggregate volume fraction well below $1/24$ (Supporting Information, section S4.3).

\section{Numerical method}\label{sec:numerics}
\begin{figure}[!htbp]
\centering
\includegraphics[width=\linewidth]{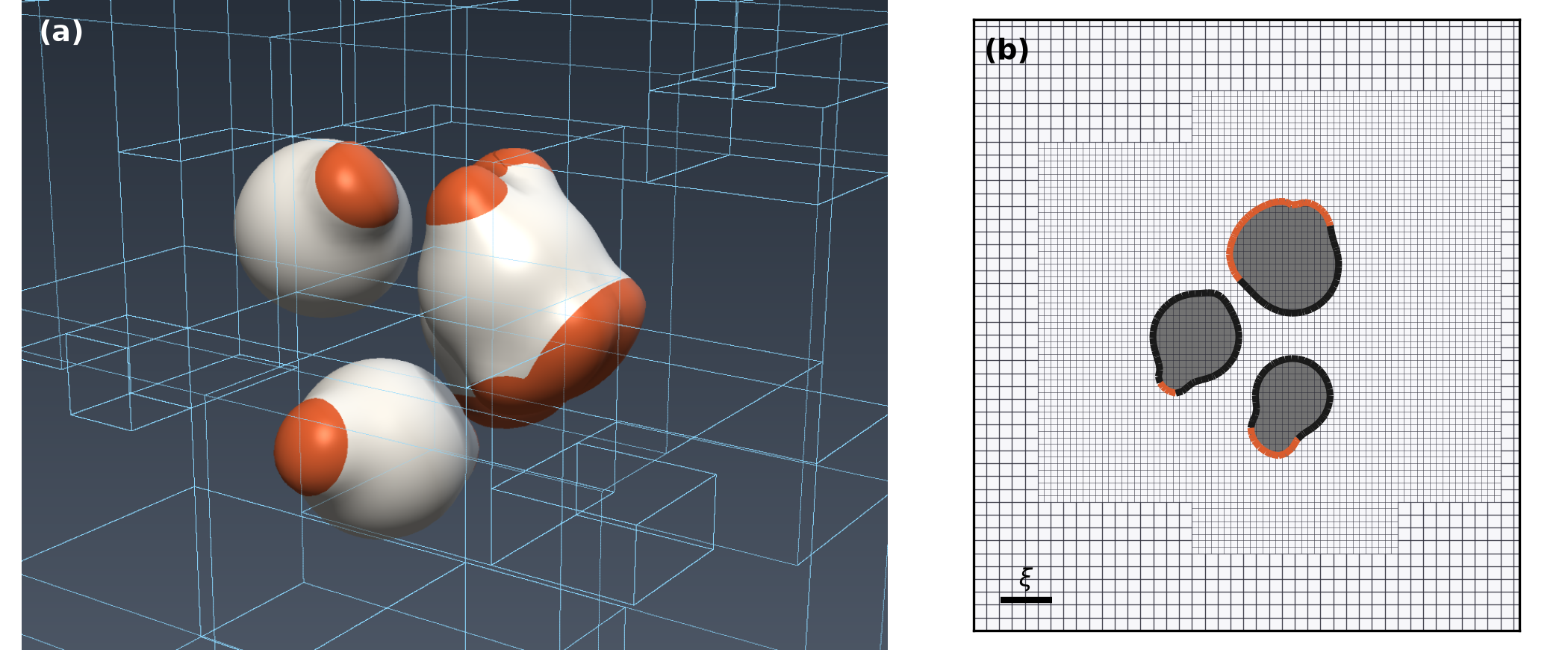}
\caption{\label{fig::amr}The adaptive grid around a growing cluster. Four nuclei, each with two sticky patches shown in colour, were placed at the corners of a tetrahedron. Two of them have merged, so three aggregates are visible. (a) The zero level set inside the finest refinement boxes. (b) A slice showing the cells of all three levels. The bar is one $\xi$.}
\end{figure}

A scalar field $\psi(t,\mathbf{x})$ defines the geometry through its sign, with $\Omega^\pm=\{\mathbf{x}:\pm\psi>0\}$ and $\Gamma=\{\mathbf{x}:\psi=0\}$, and an interface advancing at normal speed $v_n$ evolves by $\partial_t\psi+v_n|\nabla\psi|=0$.\cite{osher1988fronts} We store $\psi$ on a block-structured adaptive refinement hierarchy,\cite{zhang2019amrex} because a uniform grid at the required spacing does not fit in memory in three dimensions. The finest level has eight cells per $\xi$ and covers a band of six cells about $\Gamma$ (figure \ref{fig::amr}). The population computations below were instead run on a uniform grid at four cells per $\xi$, for the reason given at the end of this section. Every setting is tabulated in the Supporting Information, section S1.

In each step we solve the quasi-static diffusion problem for $\rho$ on $\Omega^+$ with an embedded-boundary finite-volume discretisation. At each point of the interface the sign of $\psi^s$ determines whether the growth condition or the zero-flux condition is imposed. The normal flux at $\Gamma$ is evaluated with a Richardson-extrapolated probe and extended off the interface as a velocity field.\cite{aslam2004partial} The level set $\psi$ is then advected with a fifth-order Hamilton--Jacobi WENO reconstruction and a strong-stability-preserving Runge--Kutta step,\cite{jiang2000weno,shu1988efficient} and it is reinitialised.\cite{russo2000remark,min2010reinitializing} The nucleation events of the step are then determined, and each inserts a new nucleus at its site. Finally the mergers of the step are carried out, and the size and shape of every aggregate are recorded. In the resolved branch the aggregates are rigid. The level set of each aggregate is redrawn from its stored shape at its accumulated displacement rather than advected, because advection erodes the aggregate. Contact is tested by counting site pairs within three cells, a range the grid resolves, whereas the bond rate and the availability are evaluated at the physical bonding range $a_W=0.15\,\xi$. Every population statistic below is an ensemble mean over independent realisations with its standard error, and every computation with the level-set solver was run on a single NVIDIA RTX A6000 GPU (48 GB).

The discretisation was verified before any physical result was computed. We used a manufactured solution on a sphere carrying two sticky patches. On the uniform grid the error converges at order $1.94$ in $\rho$ and at order $2.77$ in the interface flux when the surface is not divided into patches. With the division into patches the orders are $1.93$ and $3.38$. On the uniform grid the division into patches therefore lowers the order in $\rho$ by $0.015$ and raises the order in the flux by $0.61$. On the adaptive grid the orders are $0.98$ and $0.96$ without the division and $0.97$ and $0.96$ with it, so on neither grid does the division lower an order by more than $0.015$. By equation \ref{eq::vel} the interface velocity is a flux, so every computation on the adaptive grid is first order. The population results below were therefore computed on the uniform grid. The test of monomer conservation and four computations that verify the contact kinetics are reported in the Supporting Information, sections S8.2 and S8.4.

\section{Results}\label{sec:results}
\subsection{Numerical verification of the reduction}\label{subsec:reductionverification}
The reduction states that the monomer fraction of PAM, $m_{\mathrm{PAM}}(\theta)$, follows the solution $m_{\mathrm{LENP}}(\theta)$ of the rate equations when every coefficient of those equations is computed from the PAM configuration by equations \ref{eq::taug} and \ref{eq::kappa_derived}. No parameter is fitted, so the comparison has no adjustable quantity. For each realisation we measure the largest difference between the two curves, $\Delta=\sup_\theta|m_{\mathrm{PAM}}-m_{\mathrm{LENP}}|$, and we report its mean over one hundred realisations.

The computation imposes the hypotheses of the reduction at $<\rho(0)>=0.22$, $x=4$ and $\delta=1$, on a box of side $L_\Omega=32$ at a grid spacing of $\xi/4$, and it is continued to $\theta_{\mathrm{end}}=5\times10^{-3}$, at which $2.5$ per cent of the monomer has converted. The well-mixed parameter takes the four values $\varepsilon\in\{1,0.3,0.1,0.03\}$, which are set through the diffusivity $D$ alone. The same one hundred random seeds are used at every value of $\varepsilon$, so each seed gives four realisations that differ only in $\varepsilon$. The difference in $\Delta$ between two values of $\varepsilon$ is computed seed by seed and then averaged. Because the random variation common to two realisations with the same seed cancels in that difference, the change of $\Delta$ with $\varepsilon$ is measured with a smaller error than $\Delta$ itself.

\begin{figure}[!htbp]
\centering
\setlength{\figw}{0.62\linewidth}
\begin{tikzpicture}[font=\footnotesize]
\begin{axis}[width=0.53\figw, height=0.55\figw,
  xmode=log, ymode=log, xmin=0.02, xmax=1.6, ymin=1.0e-4, ymax=1.3e-3,
  xlabel={$\varepsilon$}, ylabel={$\Delta$},
  ylabel style={yshift=-2pt}, tick label style={font=\scriptsize}]
\addplot[only marks, mark=*, mark size=1.6pt] coordinates
  {(1,0.000791308) (0.3,0.000792498) (0.1,0.000792678) (0.03,0.000792988)};
\addplot[dashed, domain=0.13:1] {0.000791308*x};
\node[anchor=west, rotate=40] at (axis cs:0.30,3.0e-4) {\scriptsize $\propto\varepsilon$};
\node[anchor=north west] at (rel axis cs:0.05,0.97) {(a)};
\end{axis}
\begin{axis}[xshift=0.53\figw, width=0.45\figw, height=0.55\figw,
  xmode=log, ymode=log, xmin=0.045, xmax=0.27, ymin=1.0e-4, ymax=1.3e-3,
  xlabel={$<\rho(0)>$}, yticklabel=\empty,
  xtick={0.055,0.11,0.22}, xticklabels={$0.055$,$0.11$,$0.22$},
  xminorticks=false, tick label style={font=\scriptsize}]
\addplot[only marks, mark=square*, mark size=1.5pt] coordinates
  {(0.22,0.000792988) (0.155,0.00054798) (0.11,0.000425168) (0.055,0.000226001)};
\addplot[dashed, domain=0.05:0.24] {0.000792988*(x/0.22)};
\node[anchor=west, rotate=40] at (axis cs:0.10,2.3e-4) {\scriptsize slope $1$};
\node[anchor=north west] at (rel axis cs:0.05,0.97) {(b)};
\end{axis}
\end{tikzpicture}
\caption{\label{fig::reduction}Verification of the reduction. Standard errors are smaller than the symbols. (a) The discrepancy $\Delta$ against $\varepsilon$ at $<\rho(0)>=0.22$. A discrepancy of order $\varepsilon$ would follow the dashed line. (b) $\Delta$ against $<\rho(0)>$ at $\varepsilon=0.03$, which falls with exponent $0.899\pm0.002$.}
\end{figure}
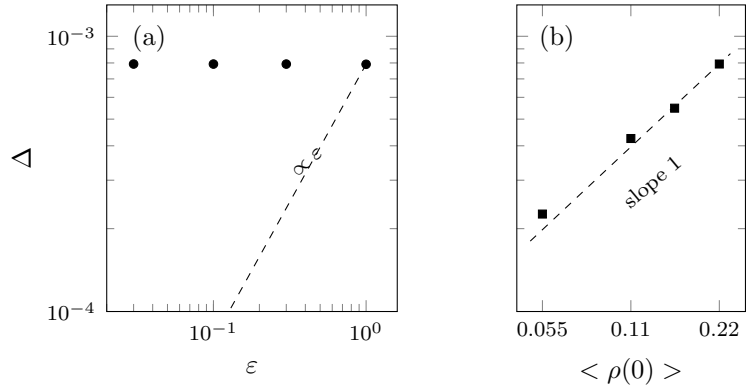

We find $\Delta=7.913$, $7.925$, $7.927$ and $7.930\times10^{-4}$ at the four values of $\varepsilon$, with a standard error of the mean near $1.4\times10^{-6}$ (figure \ref{fig::reduction}a). The standard deviation over realisations is $1.7$ per cent of $\Delta$. The two descriptions of the monomer fraction therefore differ by at most eight parts in $10^4$, which is three per cent of the monomer converted over the time analysed. The agreement is the same at $\varepsilon=1$, where the well-mixed premise is nominally violated.

The discrepancy does not depend on $\varepsilon$. Between $\varepsilon=1$ and $\varepsilon=0.03$ the paired difference in $\Delta$ is $(1.7\pm0.8)\times10^{-6}$, which is $0.2$ per cent of $\Delta$, although the measured nonuniformity of the monomer field falls seventeenfold. The fitted log-log slope of $\Delta$ against $\varepsilon$ is $-0.0006\pm0.0007$. The part of $\Delta$ that depends on $\varepsilon$ is therefore below the resolution of the computation, and the measured discrepancy comes from a different source. The computation does not resolve the order in $\varepsilon$ and so does not test it.

That source is a second approximation in the proof, which identifies the size of an aggregate with its volume. The two differ at relative order $<\rho(0)>$ because the advancing interface encloses free monomer. To test this attribution we repeated the computation at $\varepsilon=0.03$ with $<\rho(0)>$ varied fourfold. This second series was added after the first result. The discrepancy changes by a factor of $3.5$ and follows $\Delta\propto<\rho(0)>^{0.899\pm0.002}$ (figure \ref{fig::reduction}b), whereas a contribution independent of $<\rho(0)>$ would give an exponent of zero. The monomer fraction of PAM therefore agrees with the rate equations up to a remaining difference of relative order $<\rho(0)>$, which is the form in which section~\ref{sec:reduction} states the result.

The reduction also assumes that the sticky area is the same at every size. When that hypothesis is relaxed by increasing the sticky-area exponent from $0$ to $2/3$, $\Delta$ rises monotonically, and a reference that includes the size-dependent growth coefficient implied by the exponent accounts for two thirds of that response (Supporting Information, section S8.7).

\subsection{The kernel exponent against experiment}\label{subsec:kernelexponent}
\begin{figure}[!htbp]
\centering
\includegraphics[width=\linewidth]{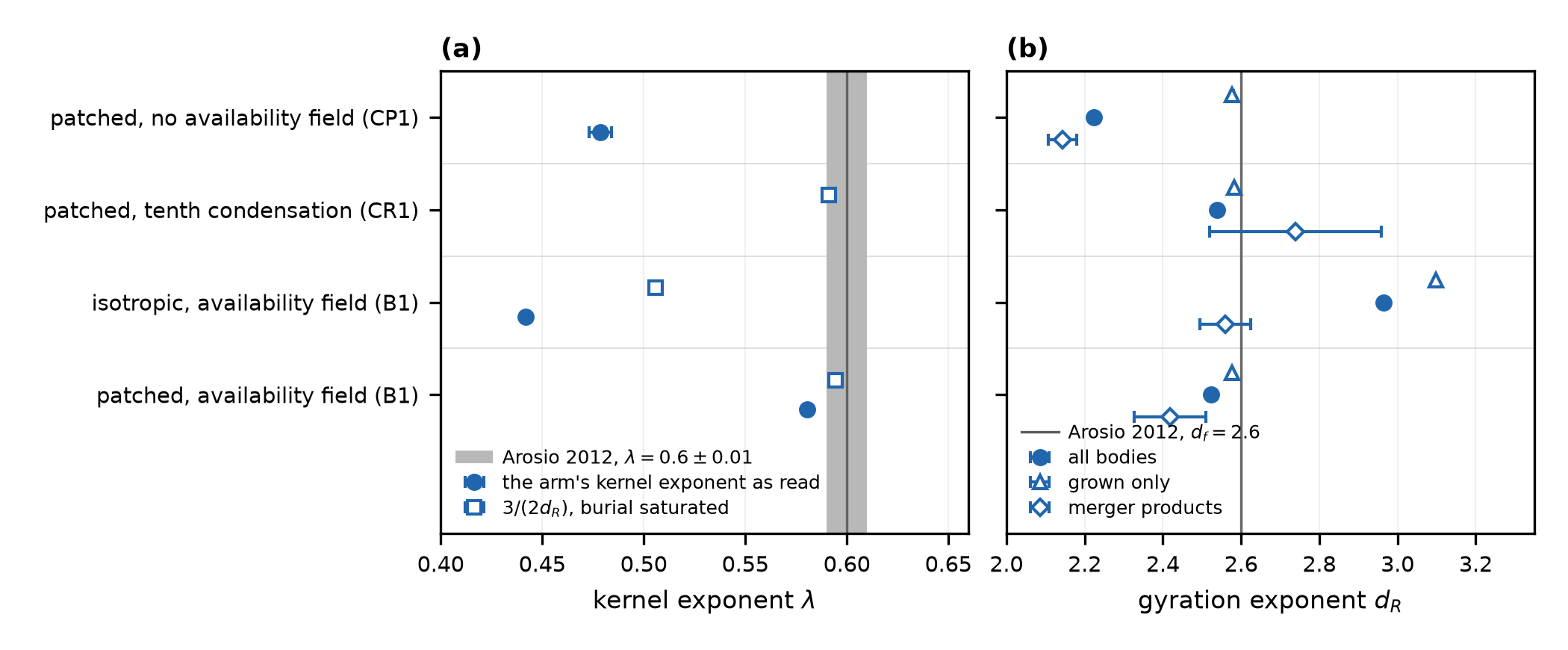}
\caption{\label{fig::arosio}Exponents of four computations against the values fitted to a monoclonal antibody.\cite{arosio2012population} (a) The kernel exponent against $\lambda=0.600\pm0.010$. Each point is described in the text. (b) The mass--radius exponent over all aggregates and, where recorded, over aggregates that grew without merging and over merger products, against the fractal dimension $2.6$.}
\end{figure}

Arosio and co-workers report $\lambda=0.600\pm0.010$ for the exponent of the product factor $(ij)^\lambda$ of the condensation kernel of a monoclonal antibody under acidic conditions, together with a fractal dimension of $2.6$ for the aggregates.\cite{arosio2012population} Equation \ref{eq::lambdakernel} predicts the kernel exponent $\lambda_\kappa$ from the availability exponent $\alpha_f$ and the mass--radius exponent $d_R$ of the computed population. We evaluated it in four computations (figure \ref{fig::arosio}). A merger product becomes compact only through growth, so $d_R$ depends on the relative rates of growth and condensation. We therefore first chose these rates in computations without the availability field, and we then added the field at the chosen rates.

Neither of the two computations without the field contains burial. The patch labels of a nucleus are written once into the region nearer to it than to any aggregate already present, so an aggregate that grows past that boundary acquires the labels of a neighbouring aggregate (Supporting Information, section S8.5). In the first of these computations the exponent was evaluated with the relation $\lambda=\alpha_\mathcal{A}-1/6$ assumed when the computation was designed, which gives $\lambda_{\mathrm{eff}}=0.4786\pm0.0055$. That relation assumes a bare area exponent of $2/3$ and a contact-distance exponent of $3$, and the same computation measures values different from both. The value is also affected by the labelling artefact, and we quote it only with both caveats.

In the second computation the condensation group $\beta_{cg}$ was divided by ten, with no other change. This raises $d_R$, evaluated over all aggregates, from $2.2237\pm0.0104$ to $2.5380\pm0.0077$, measured on $145\,156$ aggregate configurations over the recorded times. Aggregates that grew without merging remain at $2.5824\pm0.0008$. In the aged limit, where $\alpha_f$ vanishes, these relative rates give $3/(2d_R)=0.5910\pm0.0018$. That value is within one standard deviation of the published exponent, and it was obtained with one input changed and no mechanism added.

Two further computations include the field at these relative rates. In the first, the field replaces the division of the surface into sticky and non-growing patches, so the aggregates grow over their whole surface. This computation gives $d_R=2.9644\pm0.0096$ and $\lambda_\kappa=0.4417\pm0.0017$, which is far outside the published exponent. The difference is due to the growth geometry and not to burial, because a control with neither the field nor patches gives $d_R=3.0154\pm0.0114$. The second computation includes the field together with the two sticky patches. It gives $\alpha_f=-0.0140\pm0.0001$ and $d_R=2.5229\pm0.0054$ over all aggregates, and the mass--radius exponent of the aggregates that grew without merging is $2.5777\pm0.0013$. It follows that $3/(2d_R)=0.5946\pm0.0013$ and $\lambda_\kappa=0.5806\pm0.0013$, which is $1.93$ standard deviations from $0.600\pm0.010$. All three values lie inside the ranges predicted before the computation was run. Because $\alpha_f$ differs from zero by much more than its uncertainty, the population is close to the aged limit but not exactly in it, and $\lambda_\kappa$ is below the saturated value $3/(2d_R)$ by $|\alpha_f|$.

The comparison with the experiment is not exact in one respect. In our computation new nuclei form throughout the run, so small aggregates keep entering the population. In the population analysed by Arosio and co-workers, no new aggregates enter after the start. We also tested the transport-limited kernel of equation \ref{eq::kappa_derived} on its own, in a separate computation in which the aggregates do not grow and change size only by merging. In that computation we recorded every merger and compared the merger rates with the kernel. The rates increase with the ratio of the two radii, as equation \ref{eq::kappa_derived} predicts, and a kernel independent of size is excluded at between seven and twelve standard errors. The same data do not determine the value of the exponent (Supporting Information, section S8.6).

\subsection{The stability ratio against published conditions}\label{subsec:stability}
For two spheres the stability ratio of equation \ref{eq::collinskimball} reduces to
\begin{align}
    W_{i,j}=1+\mathcal{P}_{i,j}\,\Xi_{i,j},\qquad
    \mathcal{P}_{i,j}=\frac{R^3_1\big(R_i+R_j\big)}{R^2_iR^2_j},\qquad
    \Xi_{i,j}=\frac{\chi}{\eta_c\,f_if_j\,\mathrm{Da}^\infty_1} \label{eq::stabilityratio}
\end{align}
in which $\mathrm{Da}^\infty_1$ is the Damk\"ohler number at the monomer radius and the curvature-free surface rate constant (Supporting Information, section S12). We first verified this relation with the solver. For an equal pair of monomers equation \ref{eq::stabilityratio} becomes $W=1+2\Xi$. Six points from the two-body computations used to verify the contact kinetics lie on this line within $1.81$ standard errors, over a factor of sixteen in $\Xi$.

\begin{figure}[!htbp]
\centering
\includegraphics[width=\linewidth]{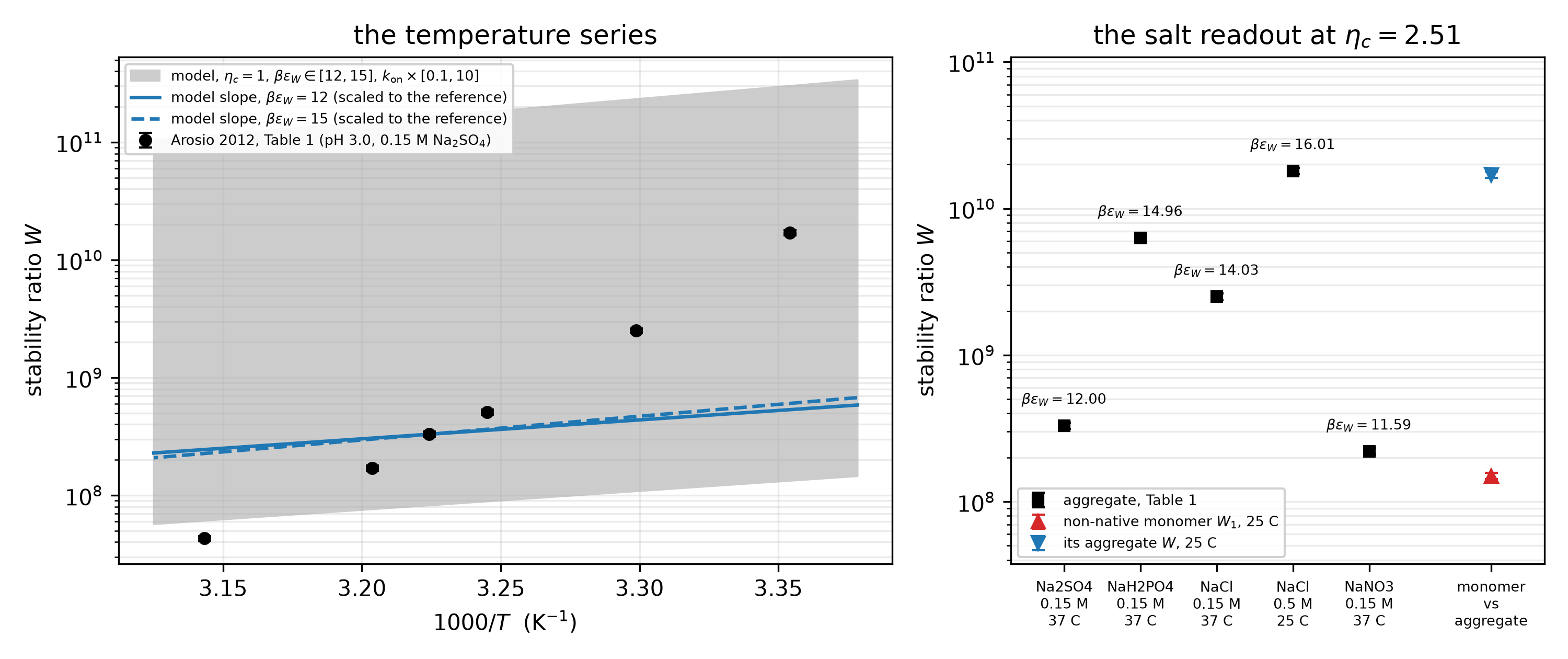}
\caption{\label{fig::w}The stability ratio of equation \ref{eq::stabilityratio} against the conditions of Arosio and co-workers.\cite{arosio2012population} \emph{Left.} The temperature series at pH $3$ in $0.15\ \mathrm{M}$ sodium sulfate. The grey band is the model at $\eta_c=1$ over the ranges of bond strength and bond rate. \emph{Right.} The salt series at the efficiency obtained from the reference condition, which shows the differences between salts rather than absolute levels.}
\end{figure}

Arosio and co-workers report fitted Fuchs stability ratios for fourteen conditions, eleven of which are for the antibody of that study at pH $3$.\cite{arosio2012population} We evaluated equation \ref{eq::stabilityratio} at each condition with the closures stated in section~\ref{subsec:availability} and compared the result with those values (figure \ref{fig::w}). The main question is whether a merger efficiency $\eta_c$ at or below unity reproduces the reference condition, which is that antibody at pH $3$ in $0.15\ \mathrm{M}$ sodium sulfate at $37\ ^\circ\mathrm{C}$, for an intrinsic bond rate inside its plausible range. Such an efficiency exists, but only on $9.1$ per cent of the grid of bond strengths and bond rates. The smallest efficiency required anywhere on that grid is $\eta_c=0.251$, at $\epsilon_W=12\,k_BT$ with the bond rate at the upper end of its range, which extends a decade on either side of the nominal value. With that bond rate thirteen of the fourteen conditions can be reached, whereas seven can be reached at the nominal rate. We discuss this dependence in section~\ref{sec:limitations}.

The temperature series is used to infer a quantity rather than to test the model. For the stability ratio, a weighted fit over the six sodium sulfate points gives an activation energy of $56.39\pm0.60\ \mathrm{kcal\,mol^{-1}}$. The slope in the model is $7.40$ to $9.24\ \mathrm{kcal\,mol^{-1}}$, which is the bond energy for $\epsilon_W$ between twelve and fifteen $k_BT$. In the model the difference between the two is the van 't Hoff enthalpy of the sticky fraction of the surface. Testing this interpretation requires an independent measurement of that enthalpy, and none is used here. The salt series is treated in the same way (Supporting Information, section S8.11).

\subsection{The two branches and the many-body enhancement}\label{subsec:twobranches}
In the resolved branch contact is tested on the level sets of the aggregates, whereas in the mean-field branch each merger is chosen at random with the rate given by a pairwise kernel. We compared the merger rate of each branch with the rate expected from two-body theory or from the rate equations.

\begin{figure}[!htbp]
\centering
\includegraphics[width=\linewidth]{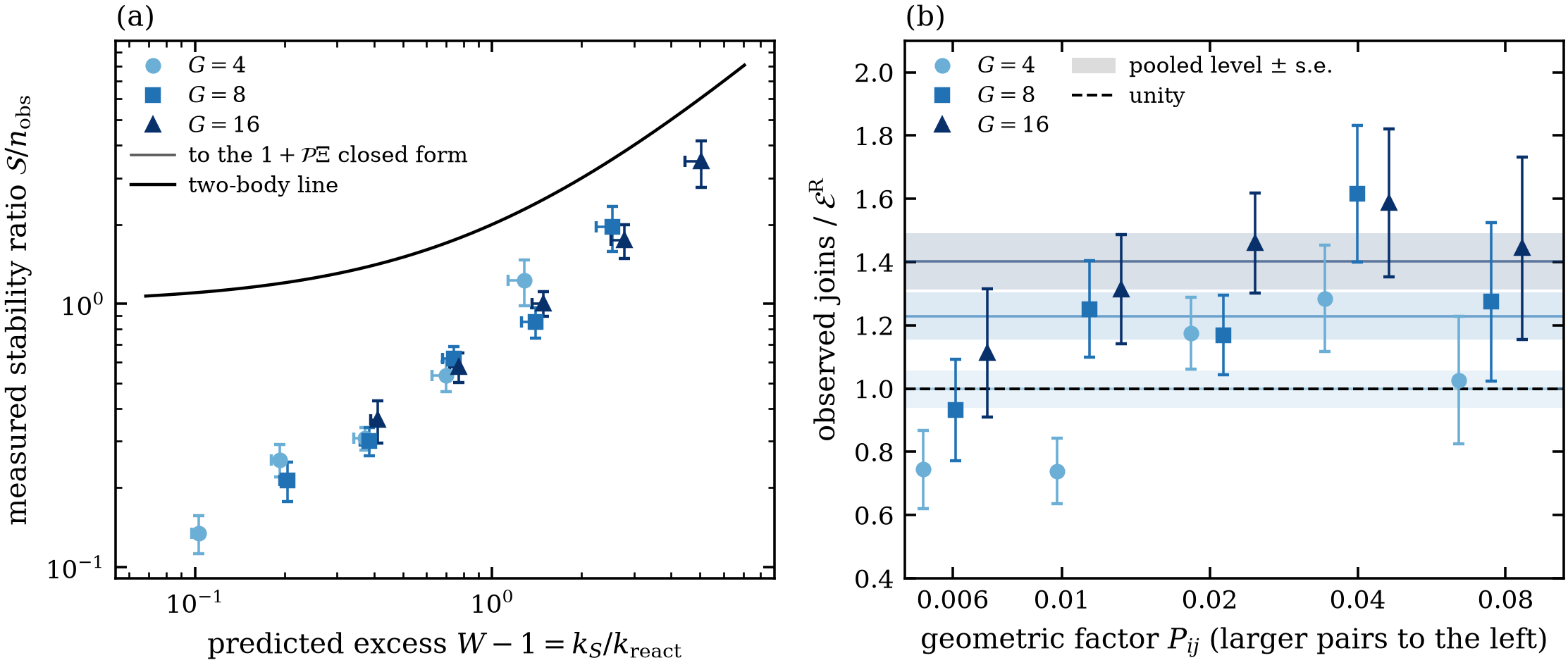}
\caption{\label{fig::resolved}Mergers in the resolved branch at three values of $G$. (a) The measured stability ratio of each size bin against the predicted $W-1$. The line is the two-body relation. (b) The ratio of observed mergers to the count $\mathcal{E}^{\mathrm R}$ expected from the reaction-limited rate, against the geometric factor $\mathcal{P}_{i,j}$ of equation \ref{eq::stabilityratio}. The bands show the ratio summed over all bins, with its error.}
\end{figure}

The resolved computation starts from a fixed set of rigid aggregates of four initial sizes, to which no new aggregates are added, at an aggregate volume fraction of $0.21875$. It was run at three values of the mobility group, $G=4$, $8$ and $16$, with four realisations each. We count every merger and compare the count with two expected counts. Each expected count is obtained by summing a pair rate coefficient over all pairs present and integrating that sum over time. The first expected count uses the two-body Collins--Kimball rate, which is the diffusion-limited rate divided by the stability ratio $W$. The second, $\mathcal{E}^{\mathrm R}$, uses the reaction-limited rate, which is the rate at which two aggregates already in contact bond. The ratio of the observed count to the first expected count is $3.824\pm0.228$, $3.109\pm0.189$ and $2.542\pm0.161$ at the three values of $G$, and it rises with the size of the pair (figure \ref{fig::resolved}a). Over the time interval that ends before any aggregate extends past half the side of the box, the same ratios are $3.185\pm0.258$, $2.691\pm0.236$ and $2.206\pm0.199$.

The excess arises because the two-body relation assumes that each aggregate is surrounded by a zone depleted of partners. At this volume fraction the capture radius exceeds the Wigner--Seitz radius, the radius of the volume available to each aggregate, for eight of the ten pairs of initial sizes. The depletion zone therefore cannot form, and dividing by $W$ understates the expected count by the factor $W/(W-1)$. When the observed count is compared with $\mathcal{E}^{\mathrm R}$ instead, the ratio no longer depends on the size of the pair. This ratio is $0.999\pm0.060$, $1.229\pm0.075$ and $1.403\pm0.089$ at the three values of $G$ (figure \ref{fig::resolved}b), which is consistent with unity only at $G=4$. The origin of the remaining excess at $G=8$ and $16$ is being tested.

The contact rule is not the source of that remaining excess. For an isolated pair, the Collins--Kimball relation predicts that the mean time before the two aggregates merge exceeds the mean time to first contact by an increment equal to the free volume divided by the reaction rate. This increment contains no diffusivity, so the resolved branch must reproduce it at every mobility. We compared the measured and predicted increments for two aggregates at $G=4$, $8$, $16$, $64$ and $256$. The difference between the measured and predicted increments, computed between realisations that share a random seed and expressed in standard errors, is $-0.28$, $+0.20$, $+1.21$, $+0.93$ and $+0.04$ at the five values of $G$. A second independent ensemble at $G=64$ gives $+1.77$. The two ensembles at $G=64$ lie on either side of the threshold of $1.6$ standard errors set for this test, and every other value is below it.

One candidate mechanism for the rise of the remaining excess with mobility is that the reaction removes pairs of aggregates that are in contact and so lowers the density of pairs at contact. We tested it in a hard-sphere computation with the same contact rule. With the reaction switched off, the contact pair density of that computation is $2.011$ at all three values of $G$, so any dependence on mobility comes from the reaction. With the reaction, the stationary contact pair density is $1.818\pm0.005$, $1.874\pm0.004$ and $1.929\pm0.004$ at the three values of $G$, whereas the two-body relation requires one half. The contact pair density implied by the many-body computation is below these values at every value of $G$. In a series of computations at increasing bond probability, the depletion produced by the reaction is at most $31.9$ per cent, whereas at $G=4$ the many-body computation is $50.4$ per cent below the reaction-free value. The mechanism therefore accounts for the direction of the remaining excess and for its mobility dependence, but not for its size.

\begin{figure}[!htbp]
\centering
\includegraphics[width=\linewidth]{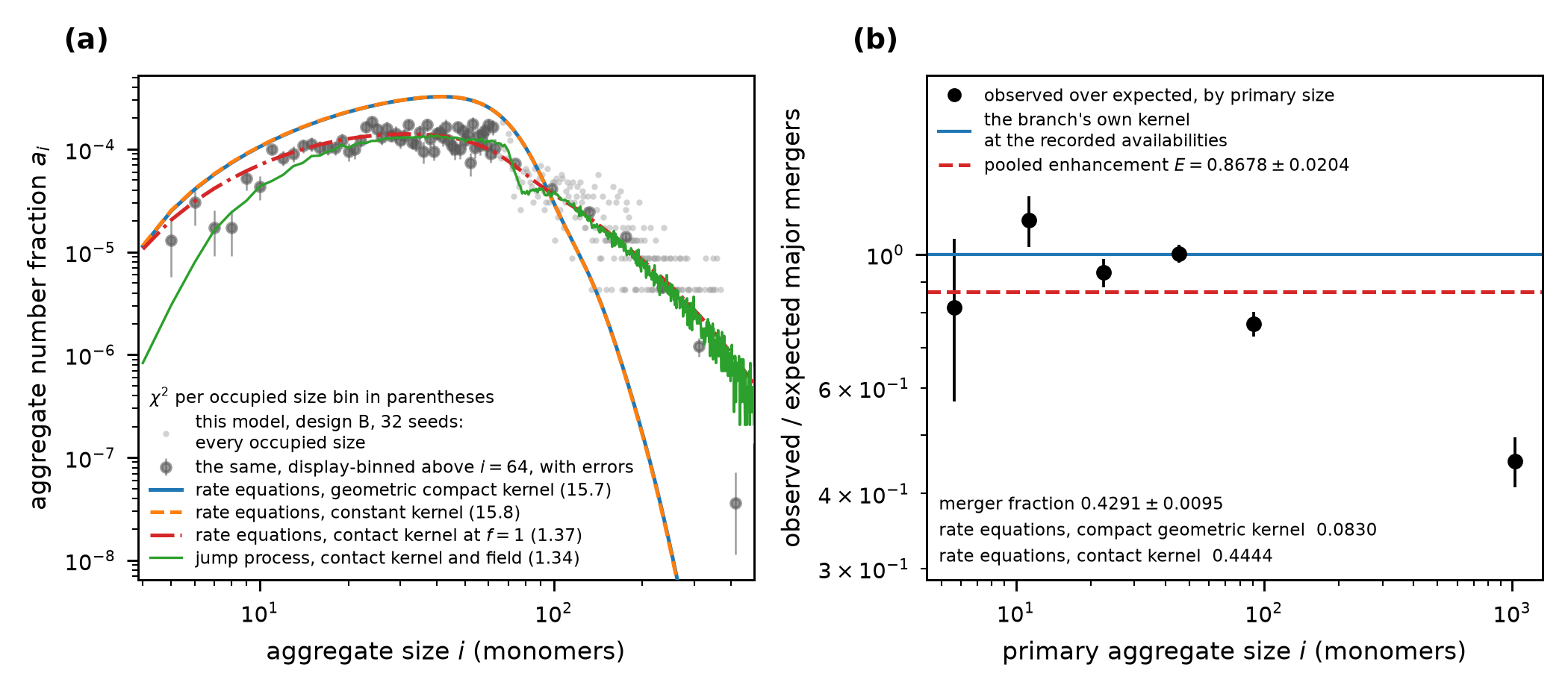}
\caption{\label{fig::census}The mean-field branch with the availability field, over thirty-two realisations. (a) The size distribution $a_i$ at $\theta_{\mathrm{end}}$ against the rate equations with the two geometric kernels and with the contact kernel, and against the finite-box stochastic simulation. The legend gives $\chi^2$ per usable bin. (b) Observed over expected major mergers by the size of the larger aggregate. The dashed line is the value over all sizes.}
\end{figure}

The mean-field computation uses the setup of the reduction verification with the mobility group raised tenfold to $G=10$, so that $\beta_{cg}=0.184$ and condensation acts. It includes the availability field at $\beta\epsilon_W=12.0$ and was continued to $\theta_{\mathrm{end}}=0.1$ over thirty-two realisations. We compared its size distribution with four references, using the $\chi^2$ per usable bin, where a bin is usable if the ensemble spread in it is nonzero (figure \ref{fig::census}a). Against the rate equations with the compact geometric kernel this statistic is $15.68$, and against the rate equations with a constant kernel it is $15.79$, so the size distribution does not distinguish these two kernels. Against the rate equations with the contact kernel from which the mergers in this computation are chosen, it is $1.367$. Against a stochastic simulation of the same rate equations in a box of the same size, which follows individual aggregates and includes the availability field, it is $1.337$. The computed population is therefore consistent with the kernel from which its mergers are chosen, whether the rate equations with that kernel are integrated or simulated.

The merger fraction, which is the share of nuclei that end in a merger, is $0.4291\pm0.0095$. The rate equations with the compact geometric kernel give $0.083$, and this factor of five comes from the form of the kernel. The same rate equations with the contact kernel give $0.4444$, so the computed fraction is $3.4$ per cent below the value expected from that kernel. For major mergers, in which the smaller aggregate is at least a quarter of the larger, the ratio of the observed count to the count expected from that kernel at the recorded availabilities is $E=0.8678\pm0.0204$. The deficit is confined to the pairs whose larger aggregate falls in the largest size bins (figure \ref{fig::census}b).

The two branches therefore differ as expected from their construction. A pairwise kernel contains no information on a third aggregate, so no contact enhancement can arise in the mean-field branch. In the resolved branch the merger rate exceeds the two-body relation by a factor between $2.2$ and $3.8$ across the three values of $G$ and the two time intervals.

\subsection{Morphology}\label{subsec:morphology}
\begin{figure}[!htbp]
\centering
\includegraphics[width=0.85\linewidth]{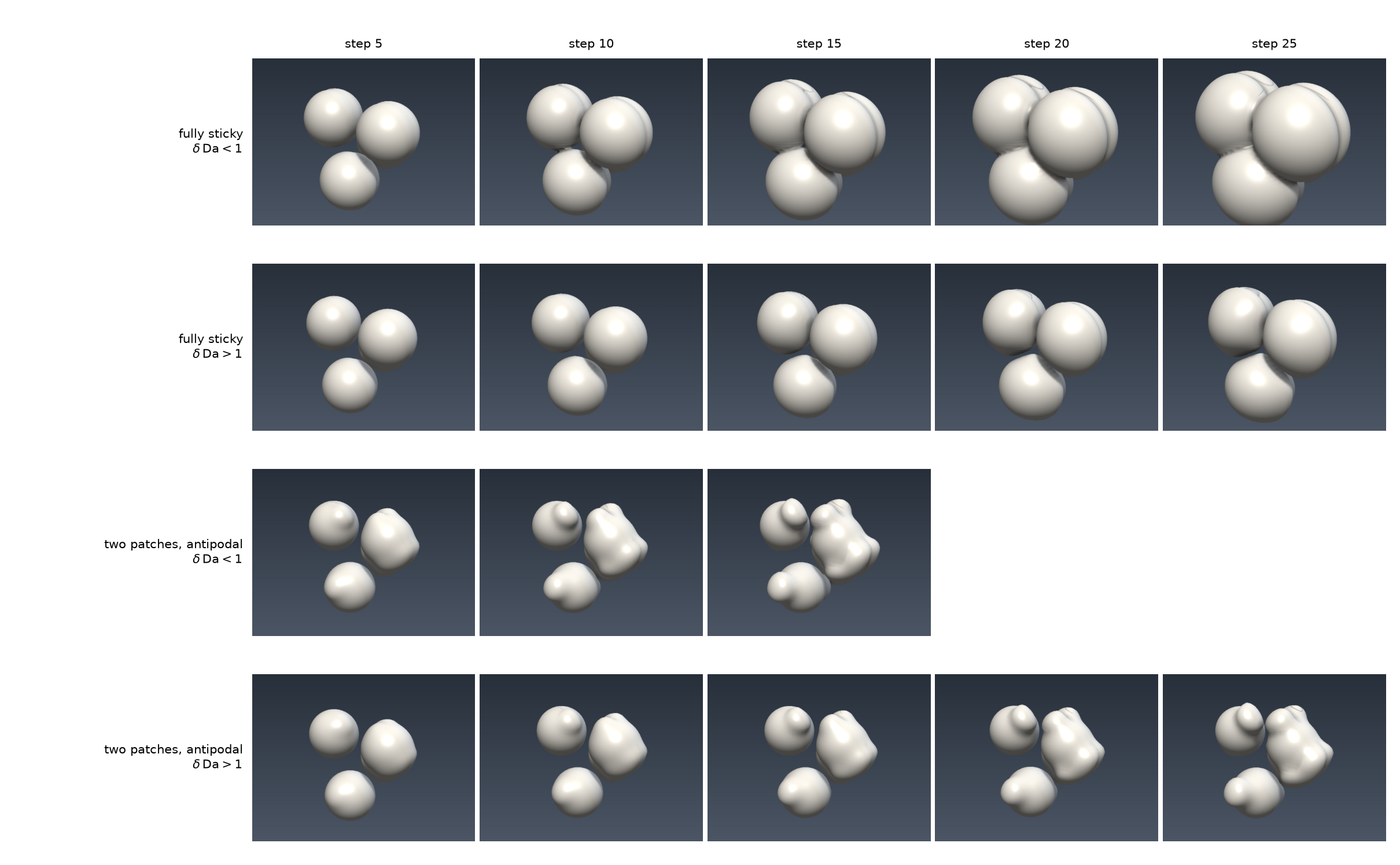}
\caption{\label{fig::gallery}The initial cluster of figure \ref{fig::amr} grown under four conditions, one per row. The surface is fully sticky in the top two rows and has two $25^\circ$ patches in the bottom two, at $\mathrm{Da}=0.25$ ($\delta\,\mathrm{Da}<1$) and $\mathrm{Da}=4$ ($\delta\,\mathrm{Da}>1$). The patched computations end when the mesh can no longer resolve the edge of a patch.}
\end{figure}

Figure \ref{fig::gallery} shows the morphologies obtained when the same initial cluster of four nuclei grows under four conditions, with a fully sticky surface or with two sticky patches, each below and above $\delta\,\mathrm{Da}=1$. The fully sticky nuclei merge into a rounded cluster, whereas the patched nuclei grow lobes at their caps. The patched cluster at $\delta\,\mathrm{Da}>1$ has the finest surface structure of the four. In that regime the finest modes that the mesh can resolve also grow, so the finest structure in that panel is at the resolution of the mesh. The dispersion relation behind the stability condition was verified degree by degree for a uniformly sticky surface (Supporting Information, section S8.12).

\begin{figure}[!htbp]
\centering
\includegraphics[width=\linewidth]{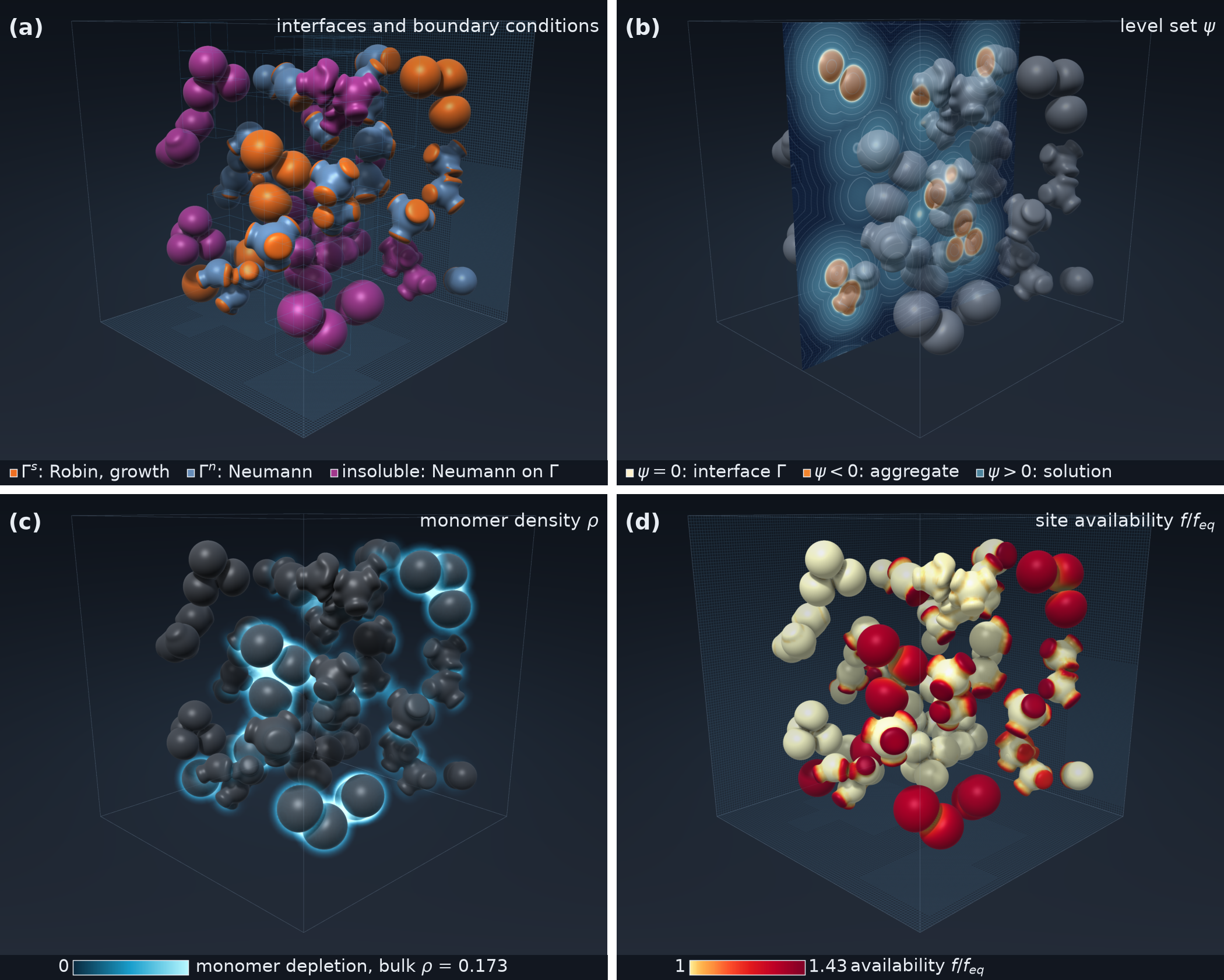}
\caption{\label{fig::views}One state ($\theta=0.068$) of a population seeded as $79$ aggregates in $33$ clusters and growing at $\mathrm{Da}=4$ in a periodic box of $L_\Omega=28$, seen by one camera. (a) The interfaces, coloured by boundary condition, with the finest refinement boxes. (b) The level set on a plane through the box, with one contour every $0.5\,\xi$. (c) The depletion of the monomer density below the bulk, integrated along each line of sight. (d) The site availability $f/f_{eq}$, integrated on the recorded surfaces with no feedback on the growth, as in figure \ref{fig::availfield}.}
\end{figure}

Figure \ref{fig::views} shows the fields of the model on one population. The population was constructed from aggregates of varied size, cluster geometry and patch arrangement in a periodic box of $L_\Omega=28$ and grown at $\mathrm{Da}=4$ (Supporting Information, section S8.13). The four panels show one frame from one camera. Panel (a) shows the interfaces, coloured by the boundary condition imposed on them in the transport solution, together with the refinement hierarchy. In panel (b) the level set is drawn on a plane through the box, so that the surfaces appear as the zero contour of one signed-distance field. Panel (c) shows the depletion of the monomer density below the bulk value, integrated along every line of sight. Panel (d) shows the site availability, integrated along the recorded surfaces as in figure \ref{fig::availfield}.

\section{Limitations}\label{sec:limitations}
The absolute level of the stability ratio depends on the intrinsic bond rate, which is known only to within about a decade. It is reproduced only with the bond rate at the upper end of that range. At the nominal bond rate, no bond strength between twelve and fifteen $k_BT$ reproduces the reference condition with a merger efficiency at or below unity, and the efficiency required there ranges from $2.51$ to $49.9$ (Supporting Information, section S8.11). An independent estimate of the bond rate would remove this dependence.

The model includes no hydrodynamic interaction between an approaching pair, which lowers the rapid-coagulation rate by a factor near two,\cite{honig1971effect} and no rotation of the aggregates.

\section{Conclusions}
We have presented a continuum free-boundary model that contains the mechanisms of nucleated polymerization and adds a spatial coordinate. In a well-mixed limit its spatial averages satisfy the rate equations term by term. With no parameter fitted, the two descriptions of the monomer fraction agree to eight parts in $10^4$. The remaining difference is of relative order $<\rho(0)>$ and does not vanish in the well-mixed limit. It arises from identifying the size of an aggregate with its volume.

The kernel exponent measured on a population with both sticky patches and the availability field is $0.5806\pm0.0013$, compared with the $0.600\pm0.010$ fitted to a monoclonal antibody. The relative rates of growth and condensation at which it was measured were reached by changing one input, the condensation group $\beta_{cg}$, and no mechanism was added. The stability ratio of the contact kinetics is consistent with thirteen of the fourteen published conditions at a merger efficiency at or below unity, but only with the intrinsic bond rate at the upper end of its range, a decade above its nominal value. In a many-body box the rigid aggregates of the resolved branch merge at between $2.2$ and $3.8$ times the two-body rate of Collins and Kimball. That excess is a many-body effect, and a pairwise kernel cannot contain it. At the lowest mobility its size is accounted for by the contact pair density of the population. At the two higher mobilities a residual remains against the reaction-limited rate, and it rises with mobility. The mechanism proposed for that residual accounts for its direction but not for its size.

Every level-set computation reported here was run on a single NVIDIA RTX A6000 GPU. Following populations through the whole growth phase will require a distributed-memory computation. In the computations reported here, growth and Brownian translation are studied separately. Computing them together at the mobility the model prescribes is the next step.

\section*{Data Availability Statement}
The data that support the findings of this study, the per-step and per-realisation records of every computation reported here, are available from the corresponding author upon reasonable request. The solver is proprietary to Examorphic, Inc. and is not distributed.

\section*{Notes}
The author declares no competing financial interest.

\begin{suppinfo}
Settings of every computation reported, the full derivations of every result stated without proof in the main text, and the computations of section~\ref{sec:results} reported at length, sections S1 to S15 (PDF).
\end{suppinfo}

\bibliography{main}

\end{document}